\documentclass[aps, prl, twocolumn,showkeys,showpacs,amsmath,amssymb,superscriptaddress]{revtex4-1}
\usepackage{graphicx} 
\usepackage{dcolumn} 
\usepackage{bm}
\usepackage{epsfig}
\usepackage{subfigure}

\begin{document}
\title{Kinetic Formulation of the Kohn-Sham Equations for {\it ab
    initio} \\ Electronic Structure Calculations}

\author{M. Mendoza} \email{mmendoza@ethz.ch} \affiliation{ ETH
  Z\"urich, Computational Physics for Engineering Materials, Institute
  for Building Materials, Schafmattstrasse 6, HIF, CH-8093 Z\"urich
  (Switzerland)}

\author{S. Succi} \email{succi@iac.cnr.it} \affiliation{Istituto per
  le Applicazioni del Calcolo C.N.R., Via dei Taurini, 19 00185, Rome
  (Italy),\\and Freiburg Institute for Advanced Studies,
  Albertstrasse, 19, D-79104, Freiburg, (Germany)}

\author{H. J. Herrmann}\email{hjherrmann@ethz.ch} \affiliation{ ETH
  Z\"urich, Computational Physics for Engineering Materials, Institute
  for Building Materials, Schafmattstrasse 6, HIF, CH-8093 Z\"urich
  (Switzerland)} \affiliation{Departamento de F\'isica, Universidade
  Federal do Cear\'a, Campus do Pici, 60455-760 Fortaleza, Cear\'a,
  (Brazil)}

\date{\today}
\begin{abstract}
  We introduce a new approach to density functional theory based on
  kinetic theory, showing that the Kohn-Sham equations can be derived
  as a macroscopic limit of a suitable Boltzmann kinetic equation in
  the limit of small mean free path versus the typical scale of
  density gradients (Chapman-Enskog expansion).  To derive the
  approach, we first write the Schr\"odinger equation as a special
  case of a Boltzmann equation for a gas of quasi-particles, with the
  potential playing the role of an external source that generates and
  destroys particles, so as to drive the system towards the ground
  state.  The ions are treated as classical particles, using the
  Born-Oppenheimer dynamics, or by imposing concurrent evolution with
  the electronic orbitals.  In order to provide quantitative support
  to our approach, we implement a discrete (lattice) model and
  compute, the exchange and correlation energies of simple atoms, and
  the geometrical configuration of the methane molecule.  Excellent
  agreement with values in the literature is found.
\end{abstract}


\maketitle

The calculation of physical properties of interacting many-body
quantum systems is one of the major challenges in chemistry and
condensed matter. In principle, this task requires the solution of the
Schr\"odinger or Dirac equations for $3N$ spatial coordinates and $N$
spin variables for electrons, where $N$ is the number of particles in
the system. Since this is computationally very intensive, the
development of approximate models to describe these systems is in
continued demand. A very successful formalism in this context is
provided by Density Functional Theory (DFT) \cite{DFTbook}, developed
by Hohenberg and Kohn \cite{hohen1} and Kohn and Sham \cite{kohn1}.
The Kohn-Sham approach to density functional theory allows an exact
description of the interacting many-particle systems in terms of a
series of effective one-particle systems, coupled through an effective
potential which depends only on the total electronic density.  In
particular, the ground state energy of the system is a functional of
the electron density, whose exact expression is however not known, due
to the complicated nature of the many-body problem.  However, over the
years, due to a tremendous amount of intensive work, many physical and
practical approximations of increasing accuracy, have continued to
appear \cite{becke1, lee1, kaxiras1, exact_XC, carpari3, review1,
  hutter, carpari2, carpari}.
\begin{figure}
  \centering
  \includegraphics[scale=0.23]{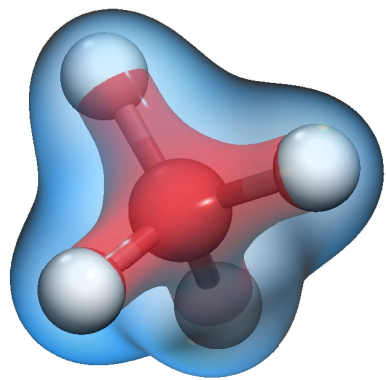}
  \caption{Methane molecule, CH$_4$. The blue and red isosurfaces
  denote low and high electronic density, respectively. Using our
  model, we have obtained for the angles between bonds, $109.47$
   degrees, and a C-H bond distance of $106$ pm.}
  \label{fig0}
\end{figure}

Kinetic theory, on the other hand, is the tool of choice for the study
of transport phenomena in dilute media.  It takes a mesoscopic point
of view by defining the probability distribution function of finding a
particle at a given position with a given momentum in the
$6$-dimensional one-particle phase-space.  This distribution function
lives at the interface between the microscopic dynamics and the
macroscopic description in terms of continuum fields representing
average quantities over microscopic degrees of
freedom~\cite{ottinger2005}.

In this Letter, we introduce a new approach to DFT, based on the
kinetic theory, whereby the Kohn-Sham orbitals are regarded as density
of quasi-particles, moving in an imaginary time, with the potential
playing the role of a source or sink of quasi-particles.  The dynamics
of these quasi-particles is governed by a kinetic equation, which,
upon Wick rotation to imaginary time, recovers the Kohn-Sham equations
in the macroscopic limit of small mean free path. Using this kinetic
approach, we compute the exchange and correlation energies of simple
atoms and molecules, particularly the methane molecule (see
Fig.~\ref{fig0}). Excellent agreement with the literature is reported.
Furthermore, due to the simple structure of the kinetic equation, it
is argued that this approach might prove valuable also for developing
an evolution equation for the total electronic density.

Let us begin by considering a many-body quantum system consisting of
$N$ electrons. In the Kohn-Sham approach to DFT, this system is
reduced to $N_o$ independent Schr\"odinger equations (Kohn-Sham
equations), subject to a mean field potential that depends only on the
total electronic density (Kohn theorem \cite{hohen1, kohn1}), namely:
\begin{equation}
 i\hbar \frac{\partial \psi_i}{\partial t} \quad =
 -\frac{\hbar^2}{2m} \nabla^2 \psi_i + V \psi_i ,
\end{equation}
where $V$ is the potential, $m$ the electron mass, and the index $i$
denotes the $i$-th electronic orbital. The total electronic density is
approximated as $\rho \simeq \sum_i^{N_o} g_i |\psi_i|^2$, where $g_i$
is the occupation number (equal to $1$ for an open-shell and $2$ for a
closed-shell) and $N_o$ denotes $N/2$ or $N/2+1$ for even and odd
numbers of electrons, respectively. Since the eigenfunctions of the
Hamiltonian form a complete basis in Hilbert space, any wave function
$\psi'$ that describes the state of a given orbital can be expanded
onto this basis,
\begin{equation}\label{expansion}
 \psi'(\vec{r},t) = \sum_n a_n \exp \left(i \frac{\epsilon_n t}{\hbar} \right) \phi_n(\vec{r}) \quad ,
\end{equation}
where $a_n$ are projection coefficients defined via the inner product,
$a_n = \langle \psi_n|\psi' \rangle$, and $\phi_n(\vec{r})$ is the
spatially dependent wave function, such that $\psi_n(\vec{r}, t) =
\phi_n(\vec{r}) T_n(t)$.  We have assumed that the potential does not
depend explicitly on time.  It is well-known that by using the Wick
rotation, which consists in replacing $t \rightarrow i\tau$ (with
$\tau$ an fictitious time), one obtains a diffusion equation with
source term,
\begin{equation}\label{diffusion}
  \frac{\partial \psi_i}{\partial \tau} = \frac{\hbar}{2m} \nabla^2 \psi_i - \frac{V}{\hbar} \psi_i \quad .
\end{equation}
Therefore, given an initial condition and using Eq.~\eqref{expansion},
we obtain $\psi_i (\vec{r},\tau) = \sum_n a_n \exp \left(-\epsilon_n
  \tau/\hbar \right) \phi_n$, where we see that for confined systems
(negative energy eigenvalues), every term in the sum grows
proportional to $\exp( |\epsilon_i| \tau/\hbar)$, while for unconfined
systems (positive energy eigenvalues) each term decreases as
$\exp(-|\epsilon_i| \tau/\hbar)$.  In both cases, after some time, the
only noticeable term in the sum will be the ground state, namely the
one that grows fastest or decreases slowest, depending on the
case. Therefore, the wave function $\psi_0$ will converge to $\psi_i
\simeq a_0 \exp (-\epsilon_0 \tau/\hbar) \phi_0$, which, upon
normalization, leads to the ground state.  This time projection
technique has been used to solve the Kohn-Sham system and obtain the
ground state for different electronic configurations \cite{wick1,
  wick2}.

However, since the diffusion equation can be derived from kinetic
theory, it must be possible to recast the Kohn-Sham equations in the
form of a kinetic equation in imaginary time.  More precisely, the
diffusion equation can be obtained from the Boltzmann equation,
\begin{equation}
  \frac{\partial f}{\partial t} + \vec{v} \cdot \nabla f = \Omega(f) + S(\vec{r}, \vec{v})
  \quad ,
\end{equation}
where $f = f(\vec{r}, \vec{v}, t)$ is a single distribution function
defined in the phase space.  Here, $S(\vec{r}, \vec{v})$ represents a
source term, and $\Omega$ is the collision operator.  For simplicity,
we approximate it with the Bhatnagar-Gross-Krook (BGK) relaxation
operator \cite{BGK}: $\Omega_{BGK} (f) = -(f - f^{eq})/\tau_K$, where
$\tau_K$ is the kinetic relaxation time, and $f^{eq}$ the equilibrium
distribution function, typically a local Maxwell distribution. Thus,
by a Chapman-Enskog expansion \cite{chapman}, for small values of the
Knudsen number $Kn$, defined as $Kn = \lambda/L$ $\lambda \propto
\tau_K$ being the mean free path and $L$ the typical system size, we
recover the diffusion equation for the density field $\rho$,
\begin{equation}\label{diffusionk}
  \frac{\partial \rho}{\partial t} = \tau_K {\cal D} \nabla^2 \rho +  {\cal S} \quad ,
\end{equation}
where $\rho = \int f \; d \vec{v} = \int f^{eq} \; d \vec{v}$, is the
zeroth order moment of the distribution function, ${\cal S} = \int S
\; d \vec{v}$ and ${\cal D}$ is defined according to the second order
moment of the equilibrium distribution, ${\cal D} = \int f^{eq}
\vec{v}^2 \; d \vec{v} / \rho$. By comparing Eqs.~\eqref{diffusion}
and \eqref{diffusionk}, we observe that the Kohn-Sham equations emerge
as the macroscopic limit of a distribution function of a gas of
quasi-particles in phase space, with the identifications $\psi_i =
\rho$, $(\hbar/2m) \psi_i = \tau_K {\cal D} \rho$, and $-V
\psi_i/\hbar = {\cal S}$.

Since we only require the zeroth, first, and second order moments of
the equilibrium distribution (they are sufficient to recover the
diffusion equation, and therefore the Kohn-Sham equations as well), we
do not need to know the exact analytical expression of the equilibrium
function (nor the intrinsic properties of the quasi-particles).
Therefore, we can expand the distribution function onto an orthogonal
basis of polynomials in velocity space up to second order.  As a
result, the distribution function, $f_i$, and the source term, $S_i$,
related to each $i$-th orbital can be written as
\begin{equation}
  f_i(\vec{r}, \vec{v}, t) = w(\vec{v}) \sum_{n=0}^{\infty} a_{n,i}^{(n)}(\vec{r}, t) H_n^{(n)}(\vec{v}) \quad ,
\end{equation}
and $S_i = - (V/\hbar) f_i$, where $w(\vec{v})$ is the weight
function, $a_{n,i}^{(n)}$ a tensor of order $n$, namely the projection
of the distribution function upon the tensorial polynomial
$H_n^{(n)}(\vec{v})$ of degree $n$,
\begin{equation}
\label{MOM}
  a_{n,i}^{(n)}(\vec{r},t)  = \int f_i(\vec{r}, \vec{v}, t)  H_n^{(n)}(\vec{v}) d^3v \quad .
\end{equation}
As mentioned before, we could choose any kind of polynomials and
weight functions that satisfy the first three order moment
constraints.  However, for simplicity we use the Hermite polynomials
with the weight $w(\vec{v}) = \exp(-\vec{v}^2/2\theta) / (2\pi
\theta)^{3/2}$, $\theta$ being a normalized temperature.  By using the
first three normalized Hermite polynomials, $H_0 = 1$, $H_1^k =
v^k/\sqrt{\theta}$, and $H_2^{kl} = (v^k v^l - \theta
\delta^{kl})/\sqrt{2}\theta$, we can readily show that the equilibrium
distribution function $f_i^{eq}$ for the $i$-th wave function becomes,
\begin{equation}\label{eq:expansion}
  f_i^{eq}(\vec{r}, \vec{v}, t) = w(\vec{v}) \psi_i(\vec{r}) \left [ 1
    + \frac{{\cal D}-\theta}{2 \theta^2} (\vec{v}^2 - 3\theta) \right ] \quad ,
\end{equation}
where ${\cal D} = \hbar/(2 m \tau_K)$.  Note that by taking $\theta =
{\cal D}$, implying that ${\cal D}$ behaves as the normalized
temperature of the gas of quasi-particles, we obtain a simpler
expression.  However, we shall keep this general form for later
applications, because, by increasing $\cal{D}$, i.e. the diffusivity
$D={\cal D} \tau_K$, the particle dynamics can be accelerated and
reach the equilibrium configuration faster.

Up to this point, there is no direct interaction between wave
functions, and therefore they all reach the same ground state.  In
order to calculate excited states, hence impose the exclusion
principle for fermions, we add an interaction potential, ${\cal V}_i$,
to the Boltzmann equation for each wave function $\psi_i$.  This
yields
\begin{equation}\label{eq:kinetic}
  \frac{\partial f_i}{\partial t} + \vec{v} \cdot \nabla f_i =
  -\frac{1}{\tau_K} \left [ \left(1 + \frac{V \tau_K}{\hbar} \right )
    f_i - f_i^{eq} \right ] - {\cal V}_i \quad ,
\end{equation}
with $ {\cal V}_i = \sum_{j<i} \Lambda_{ij} f_j$, where $ \Lambda_{ij}
= \langle \psi_i|\psi_j \rangle / \langle \psi_j|\psi_j \rangle$.  By
introducing this potential, we dynamically and sequentially remove the
contribution of excited levels $\psi_j$, with $j<i$, that do not
belong to the respective orbital $i$. This is equivalent to a dynamic
Gram-Schmidt orthogonalization procedure. Note that the interaction
potential vanishes once the system reaches the ground state, since all
orbitals are then orthogonal. The role of this potential is to
generate local quasi-particles in order to satisfy the orthogonality
conditions between the orbitals.
\begin{figure}
  \centering
  \includegraphics[scale=0.25]{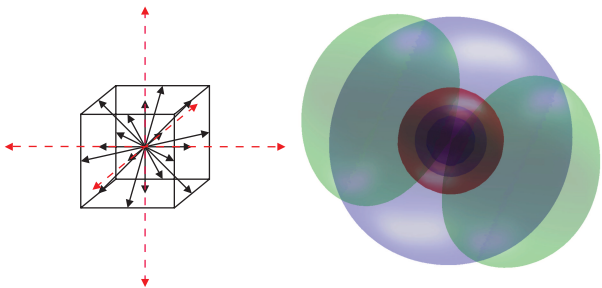}
  \caption{(Left) Lattice unit cell configuration D3Q25 ( $25$
    discrete velocities in $3$ dimensions). The solid arrows denote
    the vectors pointing to the first two neighbors, and the dashed
    lines, the ones with length $|\vec{v}_{j}| = 3$.  (Right)
    Electronic orbital density of the carbon atom, the red, blue, and
    green isosurfaces denote the first ($1s$), second ($2s$), and
    third ($2p$) orbitals at $\psi^2_{0,1,2}=10^{-5}$ au,
    respectively.}
  \label{fig1}
\end{figure}

To summarize, we have converted the system of $N_o$ original Kohn-Sham
equations into $N_o$ kinetic equations.  In this approach, we need to
solve Eq.~\eqref{eq:kinetic}, using Eq.~\eqref{eq:expansion}.  Note
that Eq.~\eqref{eq:kinetic} has no second order spatial derivatives
and therefore space and time go on the same footing.  This offers a
number of computational advantages, as we shall detail shortly.

The energy of each orbital can be calculated through the relation,
$\epsilon_i = -(\hbar/2) \partial \log(\langle \psi_i | \psi_i
\rangle) / \partial t$.

In order to check the validity of our approach, we develop a numerical
scheme and implement different simulations for simple atoms and
molecules. As it stands, Eq.~\eqref{eq:kinetic} looks computationally
over-demanding, since it lives in a six-dimensional phase
space. However, for the purpose of solving hydrodynamic problems, it
is known that velocity space can be constrained to a handful of
properly chosen discrete velocities $\vec{v}=\vec{v}_j$, where $j$
runs over a small neighborhood of any given lattice site.  This
strategy has spawned a powerful technique, known as Lattice Boltzmann
method, which has proven very successful for the simulation of a broad
variety of complex flows
\cite{Benzi1992145,chen1992recovery,succi2001lattice,
  electroLB,MUPHY}.  In the present work we employ the $25$ discrete
velocities shown in Fig.~\ref{fig1}. The result is the following set
of Lattice Kinetic Kohn-Sham equations (LKKS):
\begin{equation}
\begin{aligned}
  f_{ij} (\vec{r} &+ \vec{v}_j \delta t, t+\delta t) - f_{ij}(\vec{r},
  t) = \\ &-\frac{\delta t}{\tau_K} \biggl [\biggl ( 1 + \frac{V
    \tau_K}{\hbar} \biggr ) f_{ij}(\vec{r},t) - f_{ij}^{eq}(\vec{r},
  t) \biggr ] - {\cal V}_{ij} \quad ,
\end{aligned}
\end{equation}
where ${\cal V}_{ij} = \sum_{k<i} \Lambda_{ik} f_{kj}$. 
Further details on the discretization and the values of the numerical
parameters for the following simulations are presented in the
Supplementary Material \cite{supp}.

We first calculate the exchange and correlation energies for four
different atoms, H, Be, B, and C. For this purpose, we have used a
lattice size of $64^3$ cells. The measured values for the exchange and
correlation energies are given in Table~\ref{xcvalues}, together with
the computational time and number of iterations taken by the
discretized model. In Fig.~\ref{fig1}, we show the orbitals obtained
for the carbon atom, finding very good qualitative results.
\begin{table}
  \centering
  \begin{tabular}{|c|c|c|c|c|c|c|}\hline
    Atom & ${\cal V}_x$ & Exp.~${\cal V}_x$ & ${\cal V}_c$ & Exp.~${\cal V}_c$ & Time & Iterations\\ \hline
    H & $-0.310$ & $-0.310$ & $0$ & $0$ & $3$ min & $640$ \\ \hline
    Be & $-2.651$ & $-2.658$ & $-0.096$ & $-0.095$ & $17$ min & $3035$ \\ \hline 
    B & $-3.742$ & $-3.728$ & $-0.127$ & $-0.128$  & $21$ min & $3009$ \\ \hline
    C & $-5.084$ & $-5.032$ & $-0.173$ & $-0.161$  & $18$ min & $2968$ \\ \hline
  \end{tabular}
  \caption{Exchange-Correlation energies for H, Be, B,  and
    C. Computational time and the number of iterations performed by
    the model to reach the ground state are also shown. The expected values of
    exchange, ${\cal V}_x$, and correlation, ${\cal V}_c$, energies
    are taken from Refs.~\cite{becke1, lee1}.} \label{xcvalues}  
\end{table}

The differences between the obtained and expected exchange and
correlation energies are of the order of $1 \%$ except for the case of
the correlation energy of the carbon atom.  The large error for this
case is due to the larger spatial extension of $p$ orbital (see
Fig.~\ref{fig1}) which makes it sensitive to the boundary conditions.
The computational time of these simulations is measured when the
energy of the orbitals presents changes of less than $10^{-5} \%$
between two subsequent steps.

\begin{figure}
  \centering
  \includegraphics[scale=0.3]{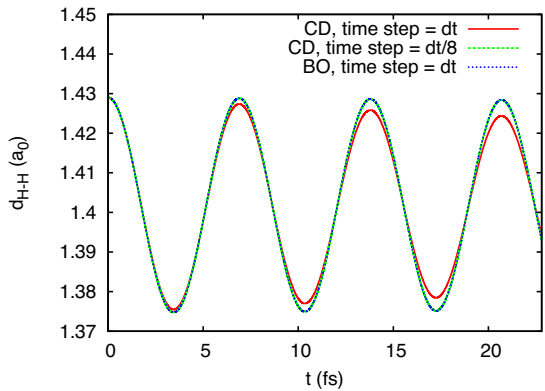}
  \caption{Vibrational mode of the diatomic molecule H$_2$. Here BO
   and CD denote Born-Oppenheimer and Concurrent Dynamics,
   respectively. The time step for this simulation is $\delta t = 0.00228$ fs.}
  \label{fig5}
\end{figure}

Our kinetic equations can also be solved in dynamic fashion, i.e. by
evolving the Kohn-Sham orbitals concurrently with the ionic motion
(Concurrent Dynamics, CD for short).  Although a rigorous theoretical
support for such a procedure remains to be developed, we performed a
study of how small the time step in the CD implementation of our
algorithm should be, in order for the system to remain sufficiently
close to the Born-Oppenhemier (BO) surface. To this purpose, we
compare both versions of our kinetic scheme, BO and CD, against each
other. For this case, we excite the H$_2$ molecule and let it vibrate
in its first mode. We have used a lattice size of $24^3$ cells. Note
that by choosing the same time step, $\delta t = 0.00228$ fs, (see
Fig.~\ref{fig5}), CD does not conserve the energy of the system, and
the amplitude of the oscillations decays in time.  However, by
choosing an eight times smaller time step, the CD shows excellent
agreement with BO, still performing three times faster than BO (CD
took $20$ minutes and BO $73$ minutes on the same machine).  The
oscillation frequency of the vibrational mode obtained by the
simulation is $4789$ cm$^{-1}$, which presents an error of $15\%$ in
comparison with the experimental value of $4161$ cm$^{-1}$
\cite{H2freq}. This discrepancy is not inherent to our approach, but
is rather due to discretization errors. Improvements of the numerical
integration of the kinetic equations will be a subject of future work.

Finally, we build the methane molecule from scratch, by using the bare
Coulomb potential.  For this simulation, we use a lattice size of
$84^3$ cells, and place the carbon atom in the center of the
simulation zone. The hydrogen atoms are located randomly in space, and
we let the system evolve to the ground configuration.  After a few
hours, we achieved the configuration shown in Fig.~\ref{fig0}. Note
that the angles are reproduced precisely, $109.47$ degrees, but the
bond distance of $106$ pm is slightly shorter than the experimental
value of $\sim 108.5$ pm \cite{ch4bond}. This bond distance
discrepancy is due to finite-size boundary effects, which compress the
molecule. This statement is supported by further simulations with
larger system sizes. The fact that we can reproduce the angles exactly
with the bare Coulomb potential, is encouraging, since the use of
pseudo-potentials requires orbital dependent parameters, which may
significantly increase the complexity of the simulation.

Summarizing, we have introduced a new approach to DFT based on
Boltzmann's kinetic theory. We have shown that the Kohn-Sham equations
can be regarded as a macroscopic limit of a first-order Boltzmann
kinetic equation. In the kinetic Kohn-Sham equations, we assume that
the ground state of a quantum system can be regarded as a gas of
quasi-particles, which are generated or destroyed by the potential,
the different wave functions interacting with each other in such a way
as to match the orthogonality constraints. This opens a new
perspective in the interpretation of multi-electron quantum systems.

The kinetic approach offers a number of computational advantages.
First, since diffusion emerges from the underlying first-order
propagation-relaxation microscopic dynamics, space and time always
appear on the same footing (no second order spatial derivatives).
This permits to advance the system in time with a time-step scaling
only linearly with the mesh resolution, rather than
quadratically. Second, since the information always travels along
straight lines, defined by the discrete velocities $\vec{v}_j$, the
streaming is computationally {\it exact} (zero round-off error).
Third, owing to its excellent amenability to parallel computing, the
present lattice kinetic formulation is expected to prove particularly
valuable for large-scale simulations.

The lattice kinetic approach has shown excellent performance and
accuracy. It can calculate satisfactorily molecular structures by
using the bare Coulomb potential, obtaining the right geometric
configuration, as it is the case for the methane molecule, without
dealing with complicated orbital dependent pseudo-potentials.
However, for large molecular structures, the inclusion of
pseudo-potentials might well become necessary.  We have also performed
computational benchmarks and compared the performance with
Car-Parrinello molecular dynamics \cite{carpari} (CPMD), finding
satisfactory results for the computational time to ground state, at
least for simple molecules.  The details of these benchmarks are
provided in the Supplementary Material \cite{supp}.

Finally, the present kinetic approach appears well positioned to solve
the evolution equation for the total electronic density, by
implementing the analogous of the Chapman-Enskog expansion for
classical fluids. This, together with improvements of the
discretization, such as local grid refinement, should permit to extend
the present lattice kinetic approach to many challenging {\it ab
  initio} calculations, possibly including those currently handled by
time-dependent density functional theory \cite{EGROSS}.

\begin{acknowledgments}
  Prof. E. Kaxiras is kindly acknowledged for many invaluable
  discussions and practical suggestions, including the one of
  performing the methane molecule calculation.  We thank Joost
  Vandevondele for discussions. We acknowledge financial support from
  the European Research Council (ERC) Advanced Grant
  319968-FlowCCS. MM would also like to acknowledge many fruitful
  discussions with Tobias Kesselring.
\end{acknowledgments}

\appendix

\section{Supplementary Material}

\section{Theoretical Background}

The aim of this supplementary material is to show the discretized form
of the kinetic equations, which in the macroscopic limit reproduce the
Kohn-Sham equations with imaginary time,
\begin{equation}\label{diffusion}
  \frac{\partial \psi_i}{\partial t} = \frac{\hbar}{2m} \nabla^2 \psi_i - \frac{V}{\hbar} \psi_i \quad .
\end{equation}

The potential energy of the system, $V$, according to the theorem of
Kohn, should only depend on the total electronic density $\rho$, and,
by using mean field approximations, contains four parts $V = V_{ext} +
V_{ee} + V_{xc}$, where $V_{ext}$ is the external potential due to
ions and/or external interactions, $V_{ee}$ is the interaction between
electrons, and $V_{xc}$ is the exchange-correlation interaction. For a
multi-electronic system interacting with $N_i$ ions, we can write
$V_{ext}$ as, $V_{ext}(\vec{r}) = -(1/4\pi \varepsilon_0) \sum_i^{N_i}
(Z_i e^2)/(|\vec{r}-\vec{R}_i|)$, where $Z_i$ is the atomic number of
the $i$-th ion, $e$ is the charge of the electron, and $R_i$ the
position of each ion. The electron-electron interaction, $V_{ee} = e
\Phi$, is obtained by solving the Poisson equation for the electric
potential $\Phi$, $\nabla^2 \Phi = e \rho /\varepsilon_0$.

For the exchange-correlation potential $V_{xc}$, we use the functional
derivative of the exchange-correlation energy ${\cal V}_{xc} = \delta
{\cal V}_{xc}/\delta \rho$.  Here we use the approach by Becke
\cite{becke1} and Lee et al. \cite{lee1}. The Laplacians and gradients
of the total density needed to calculated the exchange and correlation
potentials, are computed by using an elegant fourth-order method
recently proposed in Ref.~\cite{fourthorder}.

The ions move following classical molecular dynamics where the
Hellman-Feynman forces due to the electron density are introduced via
the electric potential $\Phi$, dictated by the spatial distribution of
the electronic charge.  In our case, the Hamiltonian of the $j$-th ion
is given by $ {\cal H}_j = \vec{P}_j^2/2 M_j + Z_j e \Phi +
\sum_{i\neq j}^{N_i} Z_i Z_j e^2 /(4\pi \varepsilon_0
|\vec{R_j}-\vec{R}_i|)$, where $Z_i$ is the atomic number, $M_j$ the
ion mass and $\vec{P}_j$ the ion momentum. Due to the Born-Oppenheimer
approximation, for each movement of the ions, the electronic density
must be updated at the ground state. Here we will not consider photon
or phonon emissions, i.e.  the electrons adapt instantaneously to the
ground state after the motion of the ions. Furthermore, for the
purpose of this work, we will only take into account the electron-ion
and ion-ion electrostatic interations, and therefore consider small
time steps in order to conserve the total energy of the system
\cite{kaxiras1}. However, to be precise, one would need to use instead
of just the electrostatic interaction, the total Hamiltonian of the
electrons. Extensions in this direction will be subject of future
research.

\section{Lattice Kinetic Description}

Our model is based on the lattice Boltzmann method. Therefore, we
define the distribution function $f_{ij}(\vec{r},t)$ that describes
the dynamics of the orbital $\psi_i$ and is associated with the
velocity vector $\vec{v}_j$. For this purpose, we will use a special
case of the Boltzmann equation ($\tau_K = \delta t = 1$) that lets the
system evolve using just the equilibrium function, \begin{equation}
  \label{LB}
  f_{ij}(\vec{r}+\vec{v}_j,t+1) = w_j \psi_i (\vec{r}, t) \left
    [ 1 + \frac{\theta - {\cal D}}{2 \theta^2}\left (3\theta - \vec{v}_j^2 \right ) \right ] \quad ,
\end{equation}
where $w_i$ are discrete weights and $\theta$ is the lattice
temperature, which are defined depending on the cell configuration,
and ${\cal D} = \hbar/m$ (note that for the analytical equilibrium
distribution in the main text, ${\cal D} = \hbar/2m$, here the factor
$1/2$ will appear as a consequence of the discretization). Standard LB
practice shows that in the continuum limit, the above equation
converges to a diffusion-reaction equation (such is the Kohn-Sham with
imaginary time, Eq.~\eqref{diffusion}) for the ``density''
$\psi_i(\vec{r},t) = \sum_j f_{ij}(\vec{r},t)$.  This is more memory
intensive than a standard finite-difference scheme, but offers
significant advantages in return. First, since diffusion is emergent
from the relaxation to local equilibrium, i.e. the rhs of (\ref{LB}),
no Laplacian is required, hence the time-step scales only linearly
with the mesh size, rather than quadratically. Second, since the local
equilibrium at the rhs is integrally shifted to the neighbor locations
$\vec{r}+\vec{v}_j$, the spatial transport component of the algorithm
is virtually exact, i.e.  zero round-off error. Third, the local
equilibrium conserves the local density to machine roundoff.  All of
the above configures LB as a very fast, second-order accurate scheme
for diffusion-reaction equation. Second order accuracy might appear
poor as compared to, say, spectral methods, but the three
aforementioned properties make the coefficients of second order errors
so small that LB has indeed been shown to provide similar accuracy as
spectral methods for, say, turbulence simulations on grids of the
order $O(10^3)$.

The functions $\psi_j$ contain a forcing term in order to satisfy the
second term in the rhs of Eq.~\eqref{diffusion} and the orthogonality
conditions between electronic orbitals. Thus, we can write the wave
function as,
\begin{equation}\label{vicious}
  \psi_j = \left (\psi'_j - V_{ar,\; j} \right )/(1+ V) \quad .
\end{equation}
Note that the second term is the interaction potential $V_{ar,\; j} =
\sum_{k<j} \psi'_k \langle \psi'_j|\psi'_k \rangle/\langle
\psi'_k|\psi'_k \rangle$ that ensures the orthogonality condition of
the wave functions $\psi_j$ and it vanishes once the system reaches
the ground state. This potential is kind of a time-dependent
Gram-Schmidt procedure. The auxiliary fields $\psi'_j$ are obtained
from the distribution functions $f_{ij}$, $\psi'_j = \sum_i^{\cal N}
f_{ij}$, where $\cal N$ is the number of velocity vectors. 

Increasing the lattice temperature, $\theta$, implies to increase the
number of velocity vectors, allowing a faster convergence but also
introducing errors of higher order derivatives. After a systematic
study of different configuration for the velocity vectors, and
following the procedure in Ref.~\cite{karli}, we have found a new cell
that gives the best performance and accuracy, D3Q25, which can be seen
in Fig. 2 in the main text. The weights $w_i$ and velocity vectors,
$\vec{v}_i$ can be derived from the equations in Ref.~\cite{karli},
for the case of $\theta = 1-\sqrt{2/5}$.

Thus, at each time step, we know the wave functions $\psi_j$, and can
calculate the electronic density using $\rho = \sum_j^{N_o} g_i
|\psi_j|^2$. For the computation of the exchange and correlation
potentials, we cannot use $\rho$, since it depends itself of the
potentials via Eq.~\eqref{vicious}, leading to highly complicated
system of non-linear algebraic equations. Therefore, as a reasonable
guess, we take the density calculated with $\psi'_j$,
$\rho'=\sum_j^{N_o} g_i |\psi'_j|^2$.

For the Poisson equation, we implement an additional diffusion
equation using $\varphi_i$ as the distribution functions that model
the electric potential $\Phi$. For this distribution, we have as
evolution equation,
\begin{equation}
  \varphi_i (\vec{r}+\vec{v}_i,t+1) = w_i \Phi \left
    [ 1 + \frac{\theta-\nu}{2\theta^2}\left (3\theta-\vec{v}_i^2 \right ) \right ] \quad ,
\end{equation}
where $\nu$ is a given diffusion parameter that tunes how fast the
potential $\Phi$ converges to the solution of the Poisson
equation. Due to the presence of charge density, we have to modify the
potential with the expression, $\Phi = \sum_i^{\cal N} \varphi_i +
\rho \nu/2\varepsilon_0$.

The energy level of each orbital can be calculated as, $\epsilon_j =
(\hbar/2) \log \left ( \langle \psi'_j| \psi'_j \rangle |_{t}/
  \langle \psi'_j| \psi'_j \rangle |_{t+1} \right)$, which
converges to a constant value once the system reaches the ground
state. For the dynamics of the ions we use the Verlet method. In order
to have an accurate value of the forcing term, we use a cubic
interpolation to calculate the gradients of the potentials at the
location of each ion. 

\section{Details of the Different Simulations}

In order to validate our model, we have performed several simulations,
whose details are presented here. First, we show the details of the
simulation for calculating the exchange and correlation energies for
four different atoms, H, Be, B, and C. Next, we implement a comparison
with Car-Parrinello molecular dynamics \cite{carpari} (CPMD) for the
case of the hydrogen molecule, and finally, we present the numerical
details of the simulation of the first vibrational mode of the
hydrogen molecule and the respective details of the calculation of the
geometry of the methane molecule.

\subsection{Exchange and Correlation energies for several atoms}

In all cases, we take a lattice size of $64^3$ cells, and fix the Bohr
radius to $6$, $9$, $10$, and $12$ cells, for H, Be, B, and C,
respectively. We have also set ${\cal D} = \nu = 0.85$. This implies
that we are fixing the relation between the Planck constant and the
electronic mass via the parameter ${\cal D}$. The consequences of this
assumption are only numerical, since we can always make a conversion
of units where the natural constants, e.g. speed of light, Boltzmann
constant, Plank constant, electronic charge and mass, etc., have
arbitrary numerical values.

\subsection{Comparison with CPMD}\label{compa}

Although a fair comparison with CPMD is not possible because it uses
pseudo-potentials, we will perform simulations for small molecules
where the valence electrons are also core electrons, i.e. hydrogen and
helium molecules. In order to model the hydrogen molecule, we consider
two hydrogen atoms separated by a distance $d = 1.401 a_0$
(equilibrium configuration), with $a_0$ the Bohr radius, and we let
the system optimize the wave function of the ground state, using both,
CPMD and our model. Since our model starts with a constant wave
function, we will use for CPMD, as initial pseudopotential and wave
function, LDA exchange-correlation term, adjusted in the CPMD input
file to optimize the wave function taking into account the
exchange-correlation model proposed by Becke-Lee-Yang-Parr (BLYP)
\cite{becke1, lee1}. The reason for this choice is to make the
comparison between both methods as fair as possible. In our model, we
have set ${\cal D} = \nu = 0.85$.
\begin{table}
  \centering
  \begin{tabular}{|c|c|c|c|c|c|c|}\hline
    & $20^3$& $24^3$ & $32^3$ & $70$ & $140$ & $280$\\ \hline
    ${\cal V}_{xc}$ & $-0.695$ & $-0.695$ & $-0.695$ & $-0.692$ & $-0.694$ &
    $-0.694$ \\ \hline
    $\%$ Error & $\ll 0.1$ & $\ll 0.1$ & $\ll 0.1$ & $0.4$ & $0.1$ &
    $0.1$ \\ \hline
    Time (s)  & $3$ & $9$ & $38$ & $13$ & $32$ & $87$\\ \hline
    Iterations &  $343$   &  $488$  & $850$ & $10$ & $10$ & $10$\\ \hline
  \end{tabular}
  \caption{Time consumption to optimize the wave function of the
    ground state for the H$_2$ molecule using our model and CPMD
    package. The errors are computed with the expected value ${\cal V}_{xc} = -0.695$
    reported in Ref.~\cite{exact_XC}. The first three columns
    correspond to LB (for lattice sizes of $20^3$, $24^3$, and
    $32^3$), and the last three to CPMD (cutoff of $70$, $140$, and
    $280$) calculations. }
  \label{time_wave}  
\end{table}

In Table~\ref{time_wave}, we show the results of the time consumption
for our model for different system sizes. For $20^3$, $24^3$ and
$32^3$, where we have chosen $a_0 = 2.5, 3.0$, and $3.6$,
respectively. We have used different lattice sizes to check
convergence and boundary effects. The lattice $20^3$ is quite small
and nevertheless can optimize properly the wave function. This is one
of the main advantages of our model, the possibility to obtain
reasonable accuracy by using few lattice sites. Also, in
Table~\ref{time_wave}, we can see the performance of CPMD for the same
configuration. Here, we have used different cutoffs for the plane wave
expansion. Note that the times are of the same order as the ones of
LB. In both cases, the tolerance of error has been adjusted to
$10^{-5} \%$.

Let us now take two atoms of hydrogen and locate them at twice their
equilibrium distance, $d = 2.8 a_0$, and let the system optimize the
position of the atoms until reaching the ground state. For our model,
we use again ${\cal D} = \nu = 0.85$, and we have chosen a time step
of $\delta t = 200$ ($0.16$ ps), and the tolerance of both, the
electronic density and ion dynamics, of $10^{-5} \%$. The lattice size
is $20^3$ cells.
\begin{table}
  \centering
  \begin{tabular}{|c|c|c|c|}\hline
    Model & LB ($20^3$) & CPMD (70) & CPMD (140)\\ \hline
    Time (s)  & 9 & 139 & 324 \\ \hline
    Final $d$ ($a_0$) & $1.402$ & $1.414$ & $1.409$\\ \hline
    Error $\%$ & $0.07$ & $0.9$ & $0.6$ \\ \hline
  \end{tabular}
  \caption{Time consumption to optimize the geometry to the ground
    states using both models, LB and CPMD. For the case of CPMD, the
    the number inside the parenthesis denotes the cutoff. The errors
    are calculated with the equilibrium distance of $d = 1.401 a_0$.}
  \label{time_geo}  
\end{table}

In the case of CPMD, we have implemented several simulations for
different cutoffs and using for the spatial optimization a tolerance
of $10^{-3} \%$ and for the wave function optimization $10^{-5}
\%$. In Table~\ref{time_geo} are shown the results of the time
consumption for both methods and also the final distance between the
atoms. For this test, CPMD has shown to be almost one order of
magnitude slower than our model, and even without achieving a very
good accuracy for the equilibrium configuration which is around $1.401
a_0$. These results show that our model can be very competitive for
{\it ab initio} simulations of the ground state. All simulations for
the comparison with CPMD have been run on a single core of an Intel
Xeon E 5430 processor at 2.66 GHz.

\subsection{Concurrence and Born-Oppenheimer Dynamics}

The numerical parameters are the same as in the previous simulations
in Sec.~\ref{compa}.

\subsection{Methane molecule}

For building the methane molecule from scratch, we use a lattice size
of $84^3$ cells, and we place the carbon atom in the center of the
simulation zone. The hydrogen atoms are located randomly in space, and
we let the system evolve towards the ground configuration. We have
chosen the same parameters as before and a Bohr radius of $11$ cells.
After $8$ hours, we have achieved the final configuration.

\bibliography{report}

\end{document}